\documentclass[sigconf]{acmart}

\usepackage{booktabs} 
\usepackage{url}
\usepackage{amsmath} 
\usepackage{graphicx} 
\usepackage{color} 
\usepackage{soul} 
\usepackage[]{todonotes} 
\usepackage{bigstrut} 
\usepackage{enumitem}
\usepackage{hyperref} 
\usepackage{listings} 
\usepackage{balance} 
\usepackage{algpseudocode} 
\usepackage{algorithm} 
\usepackage{algorithmicx} 
\usepackage{multicol}

\usepackage{amsthm} 
\usepackage{multirow} 
\usepackage{rotating} 
\usepackage{verbatim}
\usepackage{rotating} 
\usepackage{framed}
\usepackage{alltt}
\usepackage{graphicx}
\usepackage{caption}
\usepackage{lipsum}
\usepackage{color}
\usepackage{subfigure}
\usepackage{caption}
\setcopyright{none}
\acmDOI{10.1145/3283812.3283819}

\begin{document}
\title[A Fine-Grained Approach for Automated Conversion of JUnit Assertions...]{A Fine-Grained Approach for Automated Conversion of JUnit Assertions to English}

\author{Danielle Gonzalez}
\affiliation{%
  \institution{Rochester Institute of Technology}
  \state{Rochester, NY}
  \country{USA}}
\email{dng2551@rit.edu}
\author{Suzanne Prentice}
\affiliation{%
  \institution{University of South Carolina}
  \city{Columbia, SC}
  \country{USA}
}
\email{suzannep@sc.edu}
\author{Mehdi Mirakhorli}
\affiliation{%
  \institution{Rochester Institute of Technology}
  \state{Rochester, NY}
  \country{USA}}
\email{mxmvse@rit.edu}
\begin{abstract}
Converting source or unit test code to English has been shown to improve the maintainability, understandability, and analysis of software and tests. Code \textit{summarizers} identify `important' statements in the source/tests and convert them to easily understood English sentences using static analysis and NLP techniques. However, current test summarization approaches handle only a subset of the variation and customization allowed in the JUnit \texttt{assert} API (a critical component of test cases) which may affect the accuracy of conversions.
In this paper, we present our work towards improving JUnit test summarization with a detailed process for converting a total of 45 unique JUnit assertions to English, including 37 previously-unhandled variations of the \texttt{assertThat} method. This process has also been implemented and released as the AssertConvert tool. Initial evaluations have shown that this tool generates English conversions that accurately represent a wide variety of assertion statements which could be used for code summarization or other NLP analyses.
\end{abstract}
%


%
%


\maketitle
\renewcommand{\shortauthors}{Gonzalez et al.}

\section{Introduction}
\label{sec:intro}

Code is a volatile artifact that must be written, maintained, and understood by \textit{many} people with different experience and expertise. This has motivated the creation of automated code summarization tools, which analyze code, extract `important' statements, and produce natural language summaries. Through these summaries it is easier to gain a \textit{quick} understanding of what the code is \textit{actually} doing. This benefits maintenance, verification, and traceability activities. Many existing works have shown the feasibility and benefits of this approach by summarizing Java methods and classes ~\cite{moreno_automatic_2013,mcburney_automatic_2014,haiduc_supporting_2010,rastkar_generating_2011,sridhara_towards_2010}. Recently, these works have been adapted for the summarization of unit test code, and summaries of JUnit  test cases have been shown to improve the understandability of test cases~\cite{zhang2016,Panichella2016,li2016automatically,kamimura2013towards}.
While these test summarization approaches result in readable summaries, they provide heuristics for converting only a \textit{limited subset} of the complex-but-critical \texttt{assert} statement API. As a result, the applicability of these approaches for real large-scale projects is limited. In a JUnit test case, \texttt{assert} statements perform the `core action' of the test by checking a \textit{condition} after some manipulation of the code-under-test. The \texttt{assert} API provided by JUnit contains more than 45 unique combinations of conditions and parameters (expected/actual values, messages, deltas, etc.). Additional complexity arises from parameter \textit{order} not being strictly enforced. Furthermore, the popular \texttt{assertThat} method gets almost no attention in previous work, despite being one of the most complex assertions provided by JUnit (there are over 40 unique variations of this method outlined in the API). 
Due to the critical role \texttt{assert} statements play in a JUnit test case, it is important that \textit{all} possible variations are considered and properly converted to create an accurate summarization approach applicable across a large number of software projects.

This paper has two \textbf{contributions}: first, we present a detailed approach for automatically converting 45 unique variations of JUnit \texttt{assert} statements from Java to English. Our approach replicates previous test summarization techniques that already cover 8 variations of \texttt{assert} method. However, we augment this existing body of work by presenting novel heuristics for converting 37 \texttt{assertThat} variations. In addition, the entire conversion process has been implemented and released as the AssertConvert tool~\footnote{AssertConvert Tool: goo.gl/f9Z4xr}. A preliminary evaluation shows that developers find the English conversions to be accurate representations of many varieties of \texttt{assert} statements.
\vspace{-5pt}
\section{Methodology}
\label{sec:method}
To produce \textit{accurate} English conversions of \texttt{assert} methods, it is important to consider 3 key characteristics of the \texttt{assert} API~\footnote{http://junit.sourceforge.net/javadoc/org/junit/Assert.html}
:
\begin{enumerate}[leftmargin=*]
\item The `core' of an \texttt{assert} is its \textit{condition}, represented as 9 unique methods (ex: \texttt{assertEquals,assertNull,} \& \texttt{assertTrue}). 

However some methods are \textit{overwritten} multiple times to allow for different combinations of optional parameters, resulting in 36 unique combinations of \texttt{assert} conditions and parameters. \noindent\underline{Example:} all of the following are valid: \texttt{assertEquals(num1,num2)},
\texttt{assertEquals("num1 not equal to num2",num1,num2)},
\texttt{assertEquals(num1,num2,0.01)}, and
\texttt{assertEquals("num1 not equal to num2", num1,num2,0.01)}

\item Developers do not always follow the API-suggested parameter order, so extra steps must be taken to verify which parameter is which.

\item The \texttt{assertThat} method (not handled by existing approaches) accepts as parameters special methods known as `matchers' that provide more detailed conditions. Special heuristics are needed to handle these extra levels of complexity and accurately convert \texttt{assertThat} methods to English. This is demonstrated in the example below. Parameter 1 is the \textit{actual value}, and the \texttt{allOf} matcher contains 2 \textit{nested} matchers. It checks that \textit{all of} the nested matchers' conditions are met by the \textit{actual value}. \\
\noindent\underline{Example:} \texttt{assertThat("myValue", allOf(startsWith("my"), containsString("Val")))}. 
\end{enumerate}

Taking these into consideration, we use a 3-step conversion process. 
(1) Identify the \texttt{assert}'s parameters, (2) convert the \textit{expected} and \textit{actual} value parameters to English \textit{phrases}, and (3) combine these phrases into a single English sentence based on the \texttt{assert}'s \textit{condition}.
In the following subsections the heuristics our approach uses for this process are detailed. The special case of \texttt{assertThat} is discussed separately from the other types of assertion in Section~\ref{sec:that}.
\vspace{-12pt}
\subsection{Parameter Identification}
JUnit \texttt{assert}s can have up to 4 parameters, three of which are optional or not relevant in all methods. The one \textit{required} parameter is known as the \textit{actual} value, and is the result of performing some operation on the code-under-test.  The \textit{message} (displayed on assertion failure) parameter is optional but if present, it \textit{must} be the first parameter. For methods such as \texttt{assertNull} the \textit{expected} value is not needed. Otherwise, it acts as a supplement to the condition such as for \texttt{assertEquals}. Finally, there is an optional \textit{delta} parameter which is accepted when two numeric values are being compared to allow for an acceptance `margin'. Our approach to identifying parameters uses two sets of heuristics:

\noindent\underline{Part 1: Identifying Parameters by Number and Order}
The first step is to determine if the method contains \textit{optional} parameters. Each heuristic is checked \textit{in the order listed below}.

\noindent{\textbf{1 Parameter:}}
This is the easiest case, and the lone parameter is marked as the \textit{actual value}.\\ 
\noindent{\textbf{2 Parameters:}}
\begin{enumerate}[leftmargin=*]
\item{If the first parameter is a string AND the condition is \textit{Null}, \textit{Not Null}, \textit{True}, or \textit{False}}, then the first parameter is the assertion \textit{message} and the second parameter is the \textit{actual value}.
\item{Otherwise, the two parameters are the \textit{expected} \& \textit{actual values}.}
\end{enumerate}\vspace{-4px}
\noindent{\textbf{3 Parameters:}}
\begin{enumerate}[leftmargin=*]
\item{If the first parameter is a string AND the condition is \textit{Equals}, \textit{Not Equals}, \textit{Same}, \textit{Not Same}, or \textit{ArrayEquals}, then the first parameter is the \textit{message}.}
\item{If the first parameter is a string AND the condition is \textit{That}, then the first parameter is the \textit{message}.}
\end{enumerate}\vspace{-4px}
\noindent{\textbf{4 Parameters:}}
In this case, we reason that the first and fourth parameters are the \textit{message} and \textit{delta}.\\
\noindent\underline{Part 2: Distinguishing Between Values Using Type-Based Heuristics}
For all assertions with $\geq 2$ parameters (except \texttt{assertThat}, see Section~\ref{subsec:that}), the \textit{actual} and \textit{expected values} are identified using type-based heuristics adopted from Zhang et. al~\cite{zhang2016}. Any \textit{message} parameters are discarded.
\begin{enumerate}[leftmargin=*]
\item{If one parameter is a constant and the other is a method call, the constant is the \textit{expected value} and the method call is the \textit{actual value}}.

\noindent\underline{Example:} \texttt{assertEquals(myObj.getId(), 2456)}
Although the intended order is \texttt{assertEquals(expected, actual)}, the developer has switched the order, but this heuristic is still able to identify \texttt{myObj.getId()} as the \textit{actual value}.

\item{If both parameters are method calls, a check~\cite{zhang2016} is performed to see if \textit{one} was invoked by the \textit{class under test}. If so, this call is the \textit{actual value} and the other call is the \textit{expected value}}.

\noindent\underline{Example:} \texttt{assertEquals(aNum.toString(),obj.getID())} \\
The first call is a call invoked by a generic type, and the second is invoked by an object from the code-under-test. The first parameter is the \textit{expected value} and the second is the \textit{actual value}.

\item{In all other cases, it is decided that the developer followed the intended order, so the first parameter is the \textit{expected value} and the second parameter is the \textit{actual value}.}
\end{enumerate}
\vspace{-8pt}
\subsection{Condition-Based Templates for Generating Full English Sentences}
In order to properly construct English sentences, we have adopted a \textit{condition}-based template approach. The \textit{condition} is easily extracted by camel-case splitting the assertion method name and removing the term 'assert'. These rules are also adapted from previous work~\cite{Panichella2016,zhang2016}, and our novel approach for converting \texttt{assertThat} statements is described in Section~\ref{sec:that}.
\begin{enumerate}[leftmargin=*]
\item{\textbf{\texttt{assertTrue}, \texttt{assertFalse}, \texttt{assertNull}, \texttt{assertNotNull}}}
Disregarding a \textit{message} if present, there is only the \textit{actual value} parameter, and the camel-case split conditions are easily appended to the end of the \textit{actual value phrase}. 

\noindent\underline{Template:} \textit{actual value phrase} + `` is '' + \textit{condition}

\noindent\underline{Example:} \texttt{assertNotNull(myNum)} = "my num is not null".  

\item{\textbf{\texttt{assertEquals} \& \texttt{assertArrayEquals}}}
In these cases, we insert connecting terms between the two objects being compared and append the condition. 

\noindent\underline{Template:} \textit{actual value phrase} + `` and '' + \textit{expected value phrase} + ``are equal.''

\noindent\underline{Example:} \texttt{assertEquals(24,aNum)}="a num and 24 are equal". 

\item{\textbf{\texttt{assertSame} \& \texttt{assertNotSame}}}
These cases follow the same template as \texttt{assertEquals} \& \texttt{assertArrayEquals} but with small adjustments based for clarity. 

\noindent\underline{Template:} \textit{actual value phrase} +  ``[is]/[is not] identical to'' + \textit{expected value phrase}. 

\noindent\underline{Example:} \texttt{assertNotSame(24,myNum)} = "my num is not identical to 24". 
\end{enumerate}
\vspace{-5pt}
\subsection{The Special Case of \texttt{assertThat}}\label{subsec:that}
~\label{sec:that}
\begin{figure*}[h]
\hspace{-6pt}
\includegraphics[width=\textwidth,height=160px]{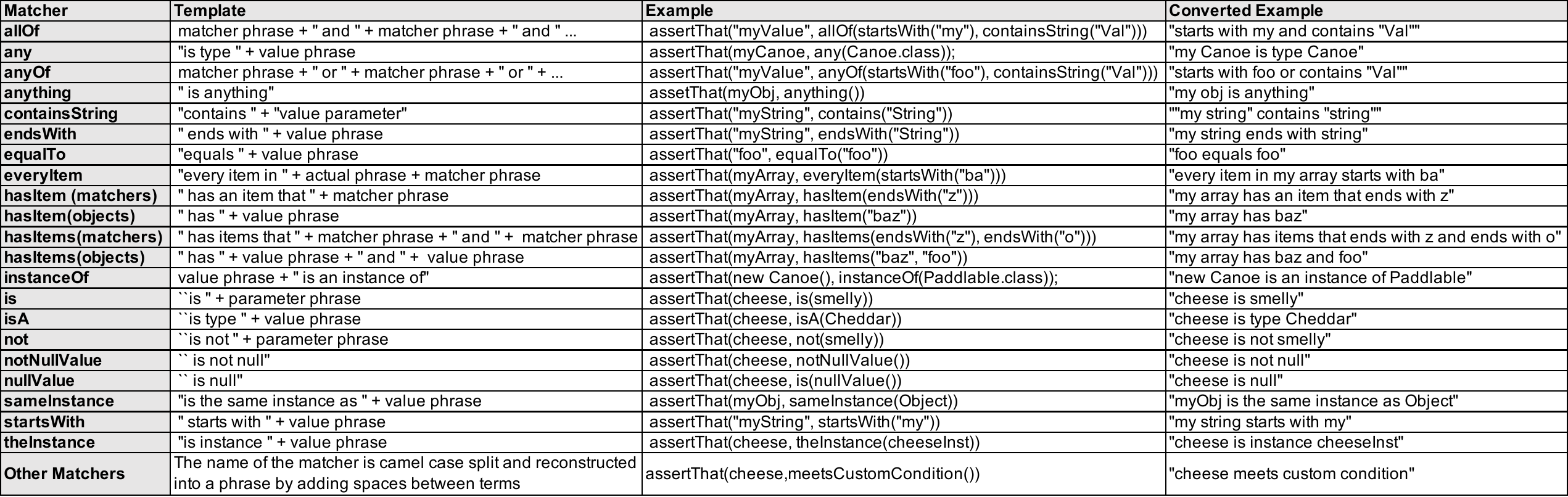}
\caption{Templates for \texttt{assertThat} Core Matchers}
\label{fig:assertThatRules}
\end{figure*}
\texttt{assertThat} is a special assertion that provides more fine-grain control over the condition being verified, even allowing multiple conditions to be checked at once. These methods can have up to three parameters; an optional `reason' (analogous to `message'), a required \textit{actual value}, and a required `matcher' method. The addition of these matcher parameters add significant complexity to parsing approaches, as many matchers are designed to accept one or more other matchers as parameters. There are over 20 `CoreMatchers' available in the Hamcrest API, plus custom matchers can be created. To demonstrate that it is possible to convert complicated assertions such as this, we have developed templates for 19 of the Hamcrest CoreMatchers (automatically recognized by JUnit) based on the API. Many matcher methods are also \textit{overloaded}, so in combination with the parameter identification and conversion processes from Sections 2.1, 2.2, and 2.4, \textbf{this approach can convert 37 variations of the assertThat method to English}. After the first round of parsing (see Section 2.1) finds the \textit{actual value}, these methods move to a special parser that extracts the \textit{condition} and \textit{expected value(s)} from the matcher(s). First, the ``top level'' \textit{condition} is extracted and stored (ex: \texttt{allOf}). Then, any nested matchers are recursively parsed and converted to a \textit{matcher phrase}. Non-matcher parameters are converted to phrases using replicated heuristics (~\ref{sec:engRules}). Finally, all phrases are combined with the "top level" condition. \\
\noindent\underline{Example:} \texttt{everyItem(startWith("My"))} becomes "every item starts with "my"". 

Figure~\ref{fig:assertThatRules} shows all the templates used to convert the matchers. For each template, if a matcher accepts \textit{only} a \textit{type or value} parameter (String, Object, etc.), the term \textit{value phrase} is used to indicate the English phrase for that parameter. If a method accepts \textit{only} matcher(s) as parameters, the term \textit{matcher phrase} is used to mean the English phrase for that matcher and its own parameters. If a matcher has been overloaded and can accept either an type/value \textit{or} matcher(s), the term \textit{parameter phrase} is used.
\vspace{-10pt}
\subsection{Converting Java Parameters to English}~\label{sec:engRules}
To convert non-matcher assertion parameters to English, we have adapted previous approaches~\cite{moreno_automatic_2013,mcburney_automatic_2014,haiduc_supporting_2010,zhang2016,rastkar_generating_2011,sridhara_towards_2010,Panichella2016,li2016automatically,kamimura2013towards}. We did not have access to the SWUM tool used by these teams to deconstruct method calls so instead we used the CoreNLP Part-of-Speech tagger~\cite{manning2014stanford}, JavaParser~\footnote{https://javaparser.org/} and SimpleNLG~\cite{gatt2009simplenlg}. This adaptation suits our purposes by adding more details to our conversions. For example, consider the assertion in Figure~\ref{fig:eval}. The second parameter is a method call. We applied the techniques from these works to identify the \textit{type} of the method caller and expand the method name. Thus, \texttt{cause.getStatusCode()} becomes "http operation failed exception status code". We also use the JavaParser's SymbolSolver to replace variable names with their type to aid in comprehension.

\vspace{-6pt}
\section{Evaluation \& Results}
\label{sec:eval}
As a preliminary check that our work towards generating understandable English representations of JUnit \texttt{assert} statements is on the right track, an evaluation was performed by 4 professional developers with JUnit testing experience.
\vspace{-8pt}
\subsection{Evaluation Setup}
The process detailed in Section~\ref{sec:method} was implemented as a tool named \textbf{AssertConvert}. 300 Java projects extracted from Github using GHTorrent~\cite{Gousi13} were given as input to the tool. From the approximately 2000 summaries generated, 25 were randomly selected. The randomly selected cases were reviewed to ensure at least 1 of each condition was included, otherwise a new random sample was reselected.
\vspace{-10pt}
\begin{figure}[!hp]
\hspace{-1pt}
\includegraphics[width=0.9\linewidth]{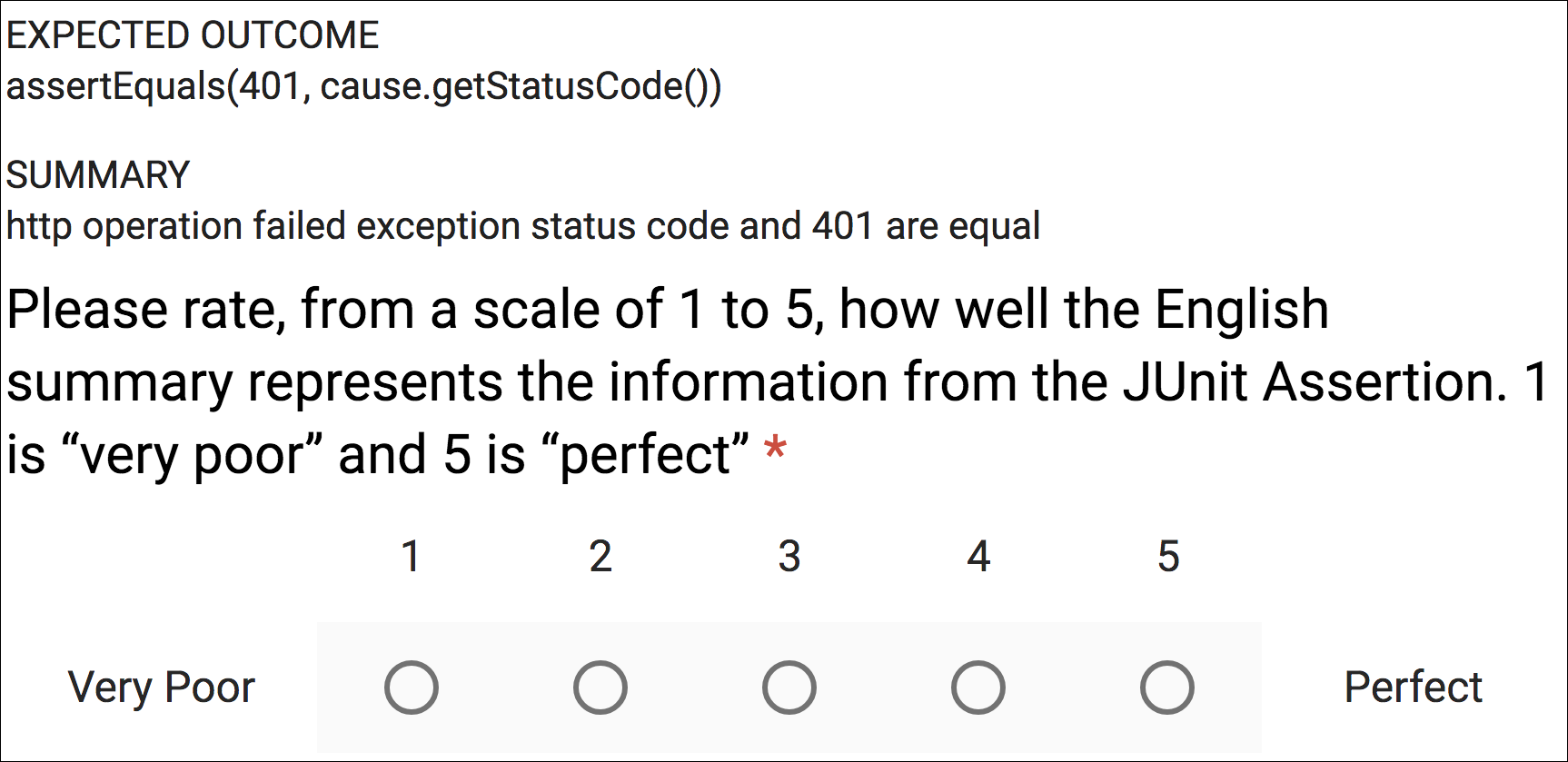}
\caption{Example from Evaluation}
\hspace{-38pt}
\label{fig:eval}
\end{figure}
\vspace{-10pt}
 For each, the original statement was presented followed by its English conversion and the participant was asked to `rate' each conversion on a scale of 1 to 5 based on how well the conversion acted as an understandable and accurate representation of each \texttt{assert} statement as shown in Figure~\ref{fig:eval}. The scale was explained as a score of 1 was 'very poor' and a 5 was 'perfect'. It was also explained to the participant that some information, particularly variable names, were \textit{added} to the summaries in order to prevent confusion. The evaluation used is available online~\footnote{https://goo.gl/forms/XafW9jx2zyAVi33V2}.

The participants in this evaluation were 4 professional developers, each with 4 or more years of experience with Java, and at least 1 year of experience regularly using JUnit for testing.

\vspace{-5pt}
\subsection{Evaluation Results}
\begin{figure}[!hp]
\hspace{-5pt}
\includegraphics[width=0.7\linewidth]{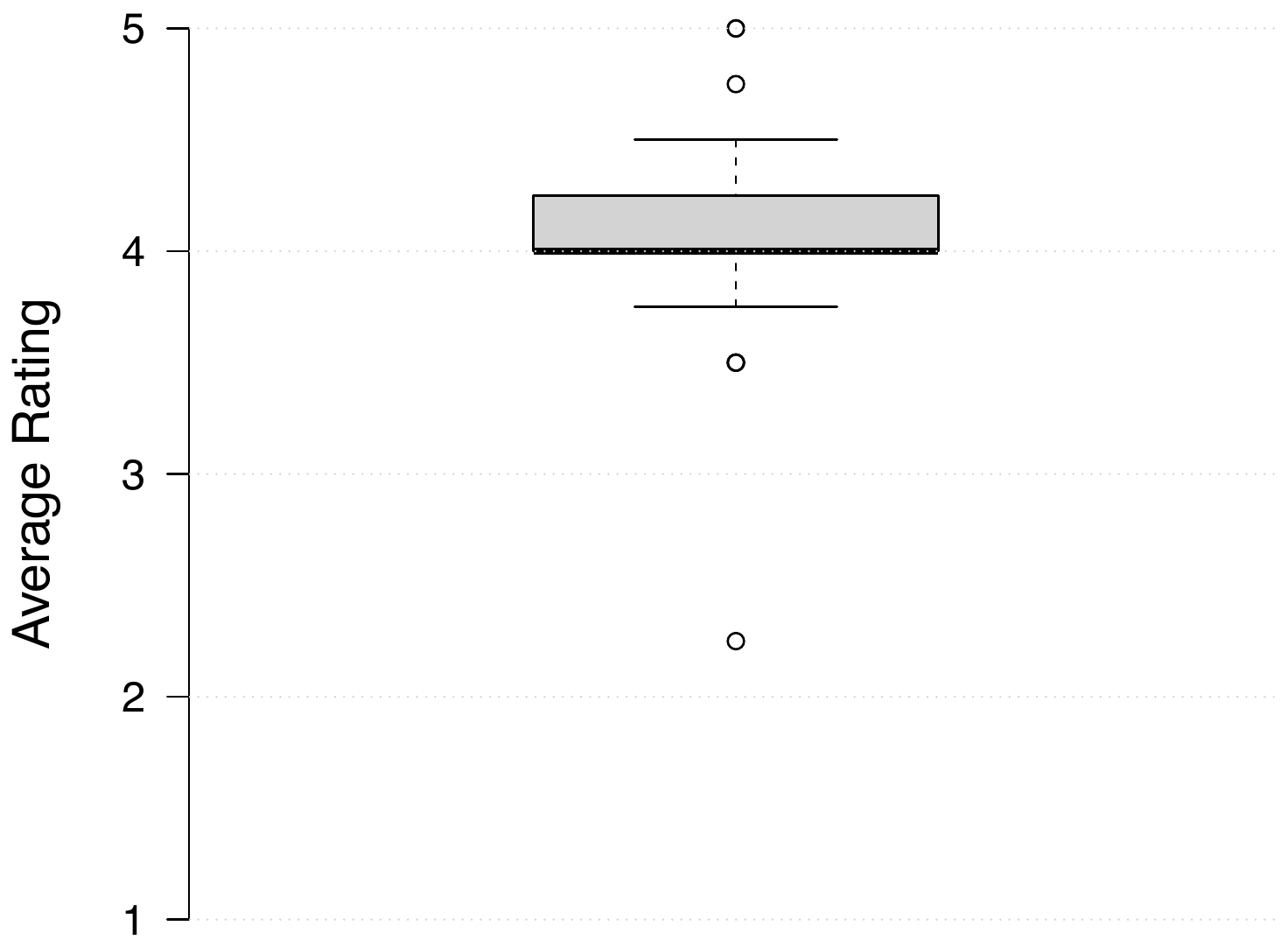}
\caption{Average ratings for the 25 evaluated conversions}
\hspace{-15pt}
\label{fig:avgs}
\end{figure}

Figure~\ref{fig:avgs} shows the range of \textit{average ratings} for the 25 conversion included in the evaluation. The lowest average was 2.26, and the highest was 5.  We also wanted to know the \textit{percentage of conversions} included in the evaluation received a \textit{majority positive} rating across the 4 participants. For a 5-point rating scale, we considered a rating of 4 or 5 to be 'positive' (3 is `neutral', and anything lower is 'negative'). Therefore, \textit{majority positive} means that a conversion received \textit{at least 3} 'positive' ratings. Table~\ref{tab:rates} reports the percentage of conversions in the evaluation which received 0 to 4 positive ratings. This shows that \textbf{84\% (21) of the conversions received a \textit{majority positive} rating from the participants}. 
\begin{table}[!h]
\scriptsize
\centering
\caption{Rating Frequency for Quality of Assert Conversions}
\begin{tabular}{|l|l|}
\hline
\textbf{\# of Positive Ratings} & \textbf{\% of Conversions} \\ \hline
0    & 4\%        \\ \hline
1      & 0\%        \\ \hline
2      & 12\%       \\ \hline
\textbf{3}      & \textbf{60\%}       \\ \hline
\textbf{4}  & \textbf{24\%}       \\ \hline
\end{tabular}
\normalsize
\hspace{-40pt}
\label{tab:rates}
\end{table}

This indicates that the English conversions were, in most cases, understood by the developers and accurately represented the origin \texttt{assert} statement. With minor edits to the parameter conversion system, this value can be improved in future work. 

\section{Related Work}
\label{sec:related}
Prior works produce English text that accurately conveys the purpose and actions of source/test code in an easily readable format. Moreno et al. have developed Java class-level summarization approaches~\cite{moreno_automatic_2013} while others focus on the method level such as Sridhara et al.~\cite{sridhara_towards_2010}, McBurney et al.~\cite{mcburney_automatic_2014}, Rastkar et al.~\cite{rastkar_generating_2011} and Haiduc et al.~\cite{haiduc_supporting_2010}. There is less work translating Java \textit{test} code all approaches (Zhang et al.~\cite{zhang2016}, Kamimura and Murphy~\cite{kamimura2013towards}, Li et al.~\cite{li2016automatically} and Panichella et al.~\cite{Panichella2016} adapt the template-based source code approaches, and also use similar approaches for distinguishing parameters. However, in no case is such fine-grained attention paid to the conversion of the \texttt{assert} statements, especially the special cases for \texttt{assertThat}. Thus, this paper is complimentary to the previous works, with the hope of improving the accuracy of JUnit test summarization.

\section{Conclusions}
\label{sec:conc}
In this paper, we described our fine-grained approach for automatically generating English conversions for JUnit assertion statements. We replicated assertion conversion techniques and improved them by increasing the number of statement varieties that can be accurately converted. In total, \textbf{45 \textit{unique} variations of \texttt{assert} statement are covered, including 37 variations of \texttt{assertThat} that have not been handled prior}. Preliminary evaluation shows the conversions can act as accurate representations of the \texttt{assert} statements. We feel that this comprehensive set of heuristics will increase the value and accuracy of English summaries of JUnit tests, enhancing test maintainability, and traceability. \textbf{The process has also been implemented and released
as the AssertConvert tool.} Alone or as part of a full test case summary, these English conversions of assertions reveal \underline{valuable} data about the test cases that can be easily analyzed at scale with other NLP techniques. In future work, we plan to make additional improvements to the parameter conversion heuristics, add heuristics for even more \texttt{assert} variations, and release the tool's source code so that the approach can be fully integrated into existing summarization tools. 
\vspace{-5px}
\bibliographystyle{ACM-Reference-Format}
\bibliography{sources}

\end{document}